# Investigation of electron beam induced migration of hydrogen in Mg-doped GaN using Eu as a probe

B. Mitchell,[1*]
D. Lee,[2] D. Lee,[3] A. Koizumi,[3] Y. Fujiwara,[3] V. Dierolf[1]

[1]*Lehigh University, 16 Memorial Dr. E, Bethlehem, PA, 18015, USA,*
[2]*Lawrence Livermore National Laboratory, 7000 East Ave L-413, CA 94550, USA*
[3]*Osaka University, 2-1 Yamadaoka, Suita, Osaka 565-0871, Japan*
*\*brm210@lehigh.edu*

**Abstract:** We demonstrate the use of hydrogen induced changes in the emission of isoelectric Eu ions, in Mg-doped p-type GaN, as a powerful probe to study the dynamics of hydrogen movement under electron beam irradiation. We identify, experimentally, a two-step process in the dissociation of Mg-H complexes and propose, based on density functional theory, that the presence of minority carriers and resulting charge states of the hydrogen drives this process.

## I. Introduction

The realization of an efficient nitride-based LED relies on the optimization of p-type conductivity. Magnesium is commonly used as an acceptor dopant to obtain p-type conductivity in GaN; however, the presence of hydrogen has been known to electrically passivate these acceptors [1]. This passivation has been attributed to the formations of neutral Mg-H complexes where the hydrogen is bonded interstitially to a neighboring nitrogen atom [2-4]. It is known that p-type conductivity can be achieved after thermal annealing at temperatures above 700°C or low energy electron beam irradiation LEEBI, both of which break up these complexes [6,7]. Previous luminescence and dissociation kinetics studies, which were performed on Mg-doped GaN (GaN:Mg), have been somewhat contradictory and difficult to interpret [7-14].

The reported discrepancies arise due to differences in both growth methods and conditions, as well as the precise thermal history of the sample at the time it was studied being either as-grown or thermally activated p-type GaN:Mg. However, we also point out that PL studies involving transitions from donors, acceptors and band levels tend to be difficult to interpret. Changes in the luminescence can have a variety of reasons, including the modifications of donors and acceptors and the opening of new excitation channels, which can compete with channels existing in the as-grown state. Circumventing this difficulty, we have investigated modifications of the Eu emission spectra in GaN:Mg co-doped with Eu, since changes in the emission spectra directly reflect alterations in the environment of the Eu ion [15]. To this end, in this paper we exploit Eu as a probe to track modifications induced by electron beam irradiation.

## II. Experiment

Mg and Eu co-doped GaN (GaN:Eu, Mg) layers (400nm thick) were grown by organometallic vapor phase epitaxy (OMVPE) on a sapphire substrate with an initial GaN buffer layer and capped by a 10nm GaN capping layer. No activation other than a cool down in $NH_3$ ambient was performed on the samples. The Eu and Mg concentrations were determined to be $3.0 \times 10^{19}$ and $2.4 \times 10^{19}$ cm$^{-3}$, respectively, by secondary ion-mass spectroscopy. For comparison, additional samples without Mg and with lower Mg concentrations were studied as well.

The CL measurements were performed in a JEOL6400 scanning electron microscope (SEM). The samples were mounted to two different stages, the first of which is a standard copper cold-finger capable of reaching temperatures of 30K. The second is a newly developed confocal microscope stage capable of simultaneous CL and site selective PL studies [16]. A single mode fiber with a 3.5μm core was used to collect the emission from the sample. A spot size of 35μm or raster area of roughly 80μm by 80μm was used to ensure that any modifications were evenly distributed over the collection area when using this stage. The mount for the sample is attached to a copper braid which allows for the sample to be cooled to roughly 105K. For all measurements the accelerating voltage was held at 10keV to ensure optimal penetration into the doped layer and minimal penetration into undoped buffer layer.

**III. Results**

In order to compare our observations with previously published findings, LEEBI experiments were performed at room temperature and 30K, and the changes in the luminescence close to the band edge were analyzed. Starting with an as-grown sample and using a beam current of 10nA, we tracked the changes in the CL spectra every second for 20s, after which there are no more significant variations [Fig. 1 (left)]. The peak at 3.38eV has been assigned to the near-band emission (NBE) and is not attributed to Mg due to its presence in un-doped samples [7, 9 and 22]. As a result of the LEEBI treatment, broad bands around 3.20eV and 3.26eV were seen to increase in intensity. No shift in energy was seen as the bands increased in intensity, which is in agreement with the theory that these bands can be attributed to a free electron to acceptor transition [7, 10, and 17].

At 30K, spectra were recorded every 0.5s and yielded much more detailed results [FIG. 1 (right)]. The as-grown spectra consisted of resolved peaks at 3.38eV, 3.29eV and 3.20eV as well as a broad peak at 3.12eV. The 3.29eV peak has been attributed to a zero phonon DAP transition with the 3.20eV and 3.12eV peaks being its phonon replicas. This increase was observed after LEEBI [7] and also after thermal annealing around 500°C [23]. After five seconds, the NBE peak at 3.38eV almost vanished and the other peaks nearly tripled in intensity. Simultaneously, the 3.20eV band shifted to 3.21eV. After continuing the irradiation for another 30 seconds, the peaks at 3.29eV and 3.21eV decreased, switched in relative intensities and shifted to 3.27eV and 3.20eV respectively.

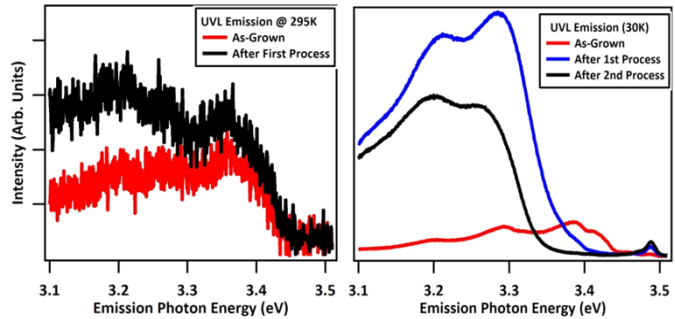

FIG. 1. (Left) UVL CL spectra taken at room temperature before irradiation (red) and after (black). An increase in a band centered 3.20eV is observed. (Right) UVL CL spectra taken at 30K, two steps are observed, one occurring after 5s (blue), and the other after 55s (black).

These observation clearly point to a two step process in the modification of the Mg-related centers at low temperatures: A first step, during which the 3.29eV peak and its replicas appear and a second step which decreases them and introduces new bands. The second step and the associated peaks have also been observed by Gelhausen et al. [18] as well and attributed to a DAP with a deeper donor, perhaps a nitrogen vacancy ($V_N$). Coleman et al. [7] observed that the modification processes take much longer and require a higher power density at room temperature compared to low. Most importantly, for the context of this paper, since our NBE observations in GaN:Eu, Mg demonstrate the same type of dissociation processes that take place in GaN:Mg, and are absent in GaN:Eu, they can be used as solid points of reference as we move to our studies to the Eu-related red emission, which we use as a probe.

Figure 2 shows the CL spectra in the Eu emission range, which contains distinct features that are only found in Mg-codoped samples, and which increase with increasing Mg concentration. This indicates that the addition of Mg introduces new Eu centers which are perturbed by the presence of Mg. In the as-grown state, one of these centers, labeled Mg/Eu1, dominates the CL emission due to its apparent superior excitation efficiency. The LEEBI procedures described above were duplicated in this region for three different temperatures and the CL response of the Eu ions was detected [Fig. 2]. We find a conversion among Mg perturbed Eu centers, which indicates, as long as we assume that neither Eu nor Mg is mobile on its own, that we are dealing with Mg-H complexes, which change their configuration under electron beam irradiation. As above, we find a two-step conversion that can be probed by the Eu-emission.

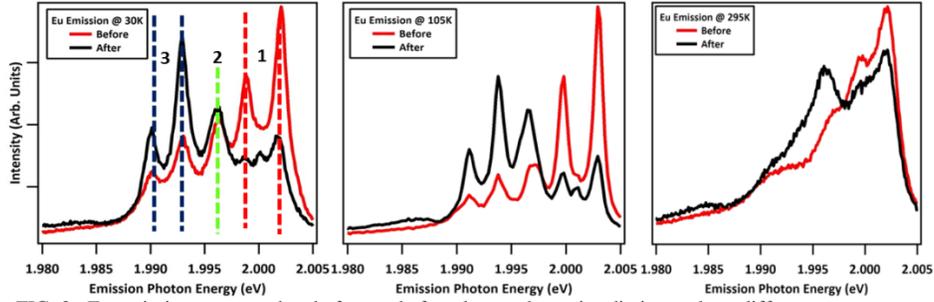

FIG. 2. Eu emission spectra taken before and after electron beam irradiation at three different temperatures.

In order to take full advantage of this, we performed site-selective combined excitation-emission spectroscopy (CEES) before and after electron beam exposure, using a beam current of 6nA at 105K [Figure 3(a) and (b)]. In this technique, the different Eu centers are resonantly excited, by tuning the excitation wavelength, and their respective emission spectra are recorded. Comparing the results to earlier studies performed on GaN:Eu [19] allows the identification of Mg-H-related centers in terms of their unique excitation and emission energies, and the determination of their relative numbers to be about 10% of the unperturbed Eu centers.

After irradiation for one minute, Mg/Eu1 is destroyed and another new site, labeled Mg/Eu2, is formed. Additionally, after irradiation, a significant increase in the center labeled Mg/Eu3 occurs. This site has the same emission profile as the majority center (Eu1) found in GaN:Eu, but has a much higher excitation efficiency than the very inefficient Eu1 center [19] under indirect excitation as can be seen by taking the intensity ratios of the CL and resonant PL spectra. More specifically, we find that both the Mg/Eu3 and the Mg/Eu2 have comparable excitation efficiency to Mg/Eu1, suggesting that the three centers share a common excitation pathway that involves the Mg-H complex.

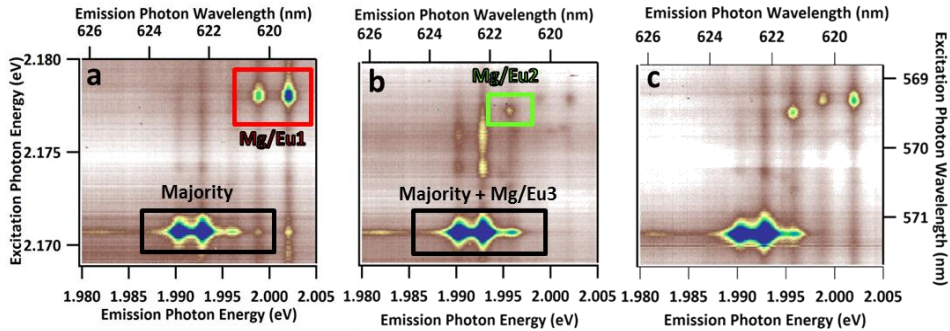

FIG. 3. CEES taken before (a) and after (b) irradiation with a beam current of 6nA. At this current the increase in Mg/Eu3 is accompanied by the appearance of a new site Mg/Eu2. A partial recovery of Mg/Eu1 and 2 is observed after heating the sample to room temperature and returning to 105K (c).

To study the thermal stability of the modification, we initially irradiated the sample at 105K, subsequently heated it to room temperature and finally cooled it back down to 105K. An interesting center modification was observed, in which the initially created Mg/Eu3 centers were significantly reduced, the Mg/Eu2 increased and the initially depleted Mg/Eu1 partially recovered. This was observed in both CL and CEES measurements shown in Fig. 3(c) and Fig. 4 (a). No such modification was detected when the sample was irradiated at room temperature and cycled to lower temperatures [Fig. 4 (b)].

The sample was also annealed in NH$_3$ at 800°C, which was the only process in which the sample reverted to its as-grown state with all of the original relative intensities restored [Fig. 4(c)]. We believe that at these temperatures, in a hydrogen rich environment, the hydrogen are recaptured into the most energetically favorable (as-grown) configurations. It should be noted that when GaN:Eu samples are annealed in an NH$_3$ ambient, no new centers are created nor changes in the luminescence properties observed [Fig. 4(d)].

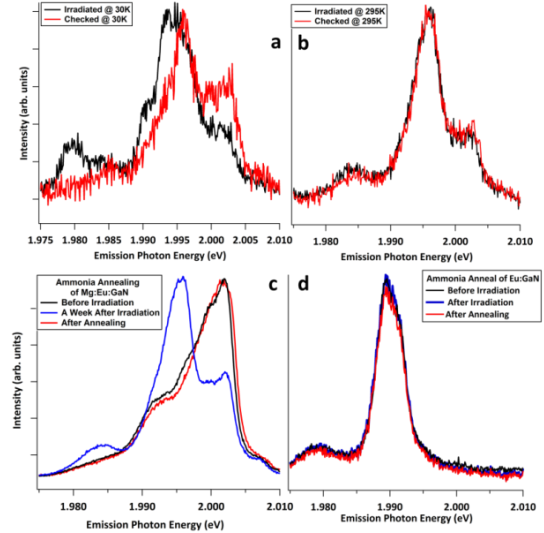

FIG. 4. The partial recovery of Mg/Eu1 and 2 is observed after irradiation is performed at 30K and the sample is heated to room temperature and brought back down to 30K (a). No recovery is seen when the sample is irradiated at 295K and cycled to 30K (b). The original spectra only returned after annealing in NH$_3$, which gives support to the H nature of the complex (c). No change is observed for either electron beam irradiation or annealing for un-doped Eu:GaN (d).

## IV. Discussion

In interpreting the structure of the Mg related Eu centers, we need to keep in mind that the observed transitions within the 4f shell of the Eu electronic configuration are sensitive mainly to the nearby environment and the displacement of the ion in regards to the lattice position of the Ga ion that it replaces. In this regard, the Mg/Eu3 center is of particular interest because no spectral shift is observed, excitation efficiency compared to the majority center Eu1. We propose that the Eu1 center consists of a Eu ion which is perturbed by a $V_N$. This configuration is quite possible, since EPR studies revealed that, the majority Er center is perturbed by a nearby $V_N$ in GaN [21]. In addition, we suggest that the Mg in the Mg/Eu3 center is sufficiently distant from the Eu ion to prevent any direct perturbation, yet close enough to positively influence the excitation efficiency.

To gain further insight on the configurations of different Mg/Eu centers and their kinetics, we performed density functional theory (DFT) calculations [25]. In these calculations, the bulk GaN system is constructed by 128 atom supercell (4x2x2 unit-cell) and Eu and Mg dopants are assumed to replace Ga site forming $Eu_{Ga}^{x}$ and $Mg_{Ga}^{'}$, while hydrogen atom goes into an interstitial site ($H_i^{\bullet}$). In order to compensate for the effect of charged defects in the system, the neutralizing background charge of the opposite sign is implicitly introduced in our calculations.

We performed rigorous searches on diverse arrangements of four different defects ($Eu_{Ga}^{x}$, $Mg_{Ga}^{'}$, $V_N^{\bullet\bullet\bullet}$ and $H_i^{\bullet}$) and the possible potential permutations for defect complexes. In a first step, we were able to confirm that the $Eu_{Ga}^{x}$-$V_N$ center (Eu1 center) is an energetically favorable configuration, see Figure 5(a). Further investigations on other defects show that $V_N$ attracts other defects and ultimately form a cluster configuration as shown in [Fig. 5(b)]. This Eu-H-N-Mg-$V_N$ configuration resembles the H-N-Mg arrangement, in which the hydrogen can capture electrons during electron beam irradiation, proposed by Meyer et al. [4]. Since the interstitial hydrogen atom strongly interacts with $Eu_{Ga}^{x}$, this configuration leads to significant modification of the $Eu_{Ga}^{x}$ position compared to the Eu-$V_N$ center. This leads us to the conclusion that this initial cluster configuration corresponds to the observed Mg/Eu1 center.

Under electron beam irradiation, and a potential change of charge of the hydrogen from $H^+$ to $H^o$, however, DFT predicts that the Mg/Eu1 is no longer the most energetically favorable configuration. With an excess electron in the system, the Mg/Eu1 become energetically less favorable, and the configuration in which hydrogen sits on the $V_N^{\cdot\cdot\cdot}$ site, is 0.17eV lower in energy, see Fig. 5(c). We correlate this arrangement to the Eu/Mg2 center since this configuration leads to an intermediate perturbation (Δ=0.065Å) of the $Eu_{Ga}^x$ center. In addition, a configuration in which the hydrogen moves away from $Eu_{Ga}^x$ toward the $Mg_{Ga}^{'}$ is equally preferable to the Mg/Eu1; see Fig. 5(d). We correlate this structure, in which the modification of $Eu_{Ga}^x$ position compared to the Eu-$V_N$ center is negligible (Δ=0.003Å), to the observed Mg/Eu3. The energetic preference of the Mg/Eu3 increases as the number of excess electrons increases, and is 0.22eV lower in energy than Mg/Eu1 when two excess electrons are present in the system.

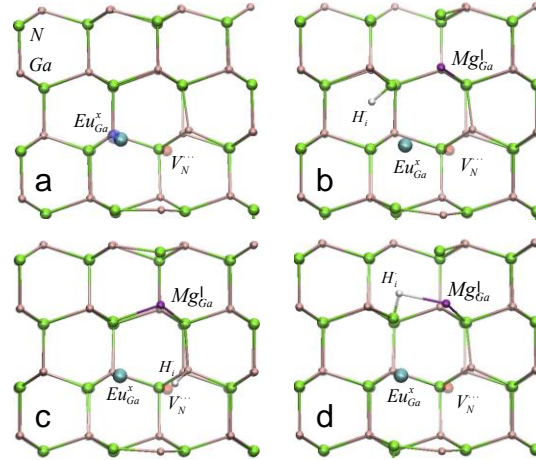

FIG. 5. DFT models for four different configurations of defect complex in Eu/Mg coped GaN: (a) A Eu-$V_N$ center which does not contain Mg or H (Eu1 center). (b) The as-grown configuration (Mg/Eu1). (c) The configuration in which the hydrogen is captured by the nearest neighbor $V_N$ after LEEBI (Mg/Eu2). (d) Metastable configuration of the hydrogen which induces a minimal perturbation on the Eu ion (Mg/Eu3).

Furthermore, our DFT calculations on the defect clusters without $Eu_{Ga}^x$, show that the Eu dopant is, in fact, a quite benign probe in GaN:Mg and that the energetics of the system are dictated by the interactions amongst the other three complex constituents. In essence, we find from the DFT calculations, three stable center configurations which exists under different charge conditions, suggesting that the addition of charge is driving the modification.

To understand the observed temperature dependence and recovery during temperature cycling, we have further performed nudged elastic band (NEB) calculations [26]. Our calculations predict that energy barrier associated with hydrogen movement from an interstitial site to a $V_N$ site is significantly higher than the movement between two interstitial sites. The energy barrier for the transition of Mg/Eu1 to Mg/Eu2 is 1.91 eV, while that to Mg/Eu3 is only 0.74 eV, with one electron added to the system. As a result, although the Mg/Eu2 is the most energetically favorable configuration under irradiation, the movement to the metastable Mg/Eu3 is more likely at lower temperatures and power densities.

Therefore, it appears that the temperature and power dependences of our system are governed by the combination of the overcoming of energy barriers and the ability of the system to hold charge. The temperature dependence of the conductivity in GaN has been studied, and it was observed that the dark current increases drastically at temperatures above 100K [27]. We suggest that the system is able to hold charge more easily, at lower temperatures, and this drives the hydrogen from Mg/Eu1 to Mg/Eu3. The metastable Eu/Mg3 will have a motivation to go back to the Eu/Mg1 by losing excess electrons or to move into the $V_N$ site by overcoming the energy barrier at the elevated temperature and thus, Mg/Eu3 is not observed after thermal cycles beginning at low temperature. At higher temperatures, the movement of the hydrogen is less efficient and so we must increase the power density to drive the modification. The combination of higher power densities and increased thermal energy gives the hydrogen the 1.91eV it requires to move to Mg/Eu2, where it is more favorable and stable. As a result, above 100K, we see a more equivalent preference for the hydrogen to move to either Mg/Eu2 or Mg/Eu3, and by room temperature the movement is predominantly into Mg/Eu2. Mg/Eu2 is a quite stable configuration and no movement to Mg/Eu1 is observed when the sample it is cycled to lower temperatures or when left at room temperature.

**V. Conclusions**

In summary, we have shown that there is a strong correlation between the changes seen in the emission from both the Eu ions and the normal emission in GaN:Mg. This correlation shows that the isoelectric nature of Eu appears to make it a good probe, and as evidenced by the results discussed herein, capably demonstrating that under LEEBI the hydrogen migrates locally within the vicinity of its as grown complex. We have verified the two distinct steps in the activation of the Mg acceptors that have previously been observed. The first involves the dissociation and charging of the Mg-H complex. The neutral hydrogen defects are then trapped locally forming $V_N$-H complexes. The second step involves the further charging and migration of the hydrogen out of the complex, yet remaining close enough to be recaptured upon heating. A pronounced temperature dependence was observed and was attributed to the reduction of activation barriers and hydrogen mobility, as well as increased conductivity at higher temperatures. Our calculations indicate that a $V_N$ is necessary for these observations to occur. The possible change in favorable Mg-H configurations and stabilities as a function of Fermi level needs to be studied.

**Acknowledgements:**
The work at Lehigh was supported by the National Science Foundation grant (ECCS- 1140038). The work at Osaka was partly supported by a Grant-in-Aid for Creative Scientific Research (Grant No. 19GS1209) and a Grant-in-Aid for Scientific Research (S) (Grant No. 24226009) from the Japan Society for the Promotion of Science. Computational work was performed under the auspices of the U.S. Department of Energy at Lawrence Livermore National Laboratory under Contract No. DE-AC52-07NA27344.

**References:**
[1] S. Nakamura, N. Iwasa, M. Senoh and T. Mukai, Jpn. J. Appl. Phys. **31**, 1258-1266 (1992).

[2] J. Neugebauer, and C. G. Van De Walle. Phys. Rev. Lett. **75**, 4452–4455 (1995).

[3] B. Clerjaud, D. Côte, A. Lebkiri, C. Naud, J. M. Baranowski, K. Pakula, D. Wasik, and T. Suski, Phys. Rev. B **61**, 8238 (2000).

[4] S. M. Myers, C. H. Seager, A. F. Wright, B. L. Vaandrager, and J. S. Nelson, J. Appl. Phys. **92**, 6630 (2002).

[5] H. Amano, M. Kito, K. Hiramatsu and I. Akasaki, Jpn. J. Appl. Phys. **28,** L2112-L2114 (1989).

[6] S. Nakamura, T. Mukai, M. Senoh and N. Iwasa, Jpn. J. Appl. Phys. **31**, L139-L142 (1992).

[7] X. Li., and J. J. Coleman, Appl. Phys. Lett. **69**, 1605 (1996).

[8] W. Götz, N. M. Johnson, J. Walker, D. P. Bour, and R. A. Street. Appl. Phys. Lett. **68**, 667 (1996).

[9] Y. Koide, D. E. Walker, B. D. White, L. J. Brillson, Masanori Murakami, S. Kamiyama, H. Amano, and I. Akasaki, J. Appl. Phys. **92**, 3657 (2002).

[10] O. Gelhausen, H. N. Klein, M. R. Phillips, and E. M. Goldys, Appl. Phys. Lett. **81**, 3747(2002).

[11] O. Gelhausen, H. N. Klein, M. R. Phillips, and E. M. Goldys, Appl. Phys. Lett. **83**, 3293 (2003).

[12] O. Gelhausen, M. R. Phillips, and E. M. Goldys, J. Phys. D **36**, 2976 (2003).

[13] O. Gelhausen, M. R. Phillips, E. M. Goldys, T. Paskova, B. Monemar, M. Strassburg, and A. Hoffmann, Phys. Rev. B **69**, 125210 (2004).

[14] A. M. Fischer, S. Srinivasan, F. A. Ponce, B. Monemar, F. Bertram, and J. Christen, Appl. Phys. Lett. **93**, 151901 (2008).

[15] D. Lee, A. Nishikawa, Y. Terai, and Y. Fujiwara, Appl. Phys. Lett. **100**, 171904 (2012).

[16] J. Poplawsky, and V. Dierolf, Microsc Microanal. **18**, 1263 (2012).

[17] M. A. Reshchikov, G.-C. Yi, and B. W. Wessels, Phys. Rev. B **59**, 13176 (1999).

[18] C. G. Van De Walle, Phys. Rev. B **56**, R10020 (1997).

[19] N. Woodward, J. Poplawsky, B. Mitchell, A. Nishikawa, Y. Fujiwara, and V. Dierolf, Appl. Phys. Lett. **98**, 011102 (2011).

[20] W. C. Dautremont-Smith, J. C. Nabity, V. Swaminathan, M. Stavola, J. Chevallier, C. W. Tu, and S. J. Pearton, Appl. Phys. Lett. **49**, 1098 (1986).

[21] A. Uedono, C. Shaoqiang, S. Jongwon, K. Ito, H. Nakamori, N. Honda, S. Tomita, K. Akimoto, H. Kudo, and S. Ishibashi, J. Appl. Phys. **103**, 104505 (2008).

[22] G. M. Criado, A. Cros, A. Cantarero, R. Dimitrov, O. Ambacher, and M. Stutzmann, J. Appl. Phys. **88**, 3470 (2000).

[23] W. Götz, N. M. Johnson, D. P. Bour, M. D. McCluskey, and E. E. HallerAppl, Phys. Lett. **69**, 3725 (1996).

[24] A. K. Viswanath, E. Shin, J. I. Lee, S. Yu, D. Kim, B. Kim, Y. Choi, and C. Hong, J. Appl. Phys. **83**, 2272 (1998).

[25] W. Kohn and L. J. Sham, Phys. Rev. **140**, A1133 (1965).

[26] H. Jónsson, G. Mills, and K. W. Jacobsen, *Nudged Elastic Band Method for Finding Minimum Energy Paths of Transitions* (World Scientific, 1998).

[27] S. J. Chung, B. Karunagaran, M. Senthilkumar, and E.-K. Suh, J. Korean Phys. Soc. **49**, 177 (2006).